\documentclass[prb,aps,twocolumn,floats,showpacs,amsmath,amssymb]{revtex4}
\usepackage{graphicx}
\usepackage{epstopdf}
\usepackage{natbib}
\usepackage{amsmath,amssymb}
\newcommand{\be}{\begin{equation}}
\newcommand{\ee}{\end{equation}}
\newcommand{\bea}{\begin{eqnarray}}
\newcommand{\eea}{\end{eqnarray}}

\begin{document}
\title{Slave Rotor Approach to Exciton Condensation in a Two-band System}

\author{Subhasree Pradhan}
\email[E-mail:]{spradhan@phy.iitkgp.ernet.in }
\author{A. Taraphder}

\affiliation{Department of Physics, Indian Institute of Technology Kharagpur, Kharagpur-721 302}
\affiliation{Center for Theoretical Studies, Indian Institute of Technology Kharagpur, Kharagpur-721 302}

\begin{abstract}

{Slave rotor mean field (SRMF) theory is employed to study exciton formation in the extended Falicov-Kimball Model (EFKM). In this theory, charge and spin (or orbital) degrees are treated as independent degrees of freedom, coupled by a constraint. Using this formalism in single as well as cluster extension, we capture the effective many body scales beyond conventional mean-field theory. While the formation of exciton is favored by the interband hybridization $V$, it is strongly influenced by the on-site Coulomb interaction $U$. Beyond a critical hybridization, there is a condensation of exciton, effectively giving rise to a crossover from metal to an excitonic insulator phase. The system goes from a metal to an excitonic insulating state, with a first order jump in the excitonic order parameter. Moreover, the behavior of excitonic averages differs from earlier results from Hartree-Fock mean-field theory. Low-$U$ results show that excitonic order parameter ($\Delta$) is continuous across the transition both for single as well as two-site approximation. The transition changes to weakly first order in the intermediate $U$ for cluster case. On the other hand, large $U$ limit shows a continuous transition for cluster but remains first order in the single-site approximation. The slave rotor also indicates an excitonic metallic region in both single and cluster cases, while there is an orbital liquid in the insulating regime in the cluster theory. The $\Delta-V$ graph shows step-like behavior when both the bands have the same parity. For cluster approximation, a second order to first order transition in $\Delta-V$ is obtained by tuning the hopping parameter of the localized band.}

\end{abstract}

\date{\today}
\maketitle

\section{Introduction}
The unabated interest in the formation and condensation of excitons~\cite{kohn, halperin-rice}, without waning in over half a century now, is primarily due to their application potential in the one hand and a deep connection to physics of charge fluctuations on the other. From the technologically useful excitonic states in semiconductors~\cite{cardona} to the charge density order via excitonic fluctutions in transition metal dichalcogenides~\cite{sk}, novel superconductors~\cite{Bi}, electronic ferroelectricity~\cite{sham,portengen, pradhan1} and so on, the physics of electron-hole bound states have captivated the interest of the scientific community for half a century and still remained alive and engrossing. 

The Coulomb attraction between the conduction band electrons and the valence band holes, under certain conditions, causes the formation of bound states, the excitons. The excitonic bound state is found in low carrier density materials with a small, direct or indirect, band gap. A plethora of theoretical and experimental work has been performed for decades to explore the physics of the excitons in a variety of systems~\cite{kohn, Wachter1, High, combescot}. 

The exciton condensation in semiconductor bilayers employs the strategy of hosting electrons in the first layer and holes in the second layer by means of electric gates which allow separate contacts to the layers. The spacer between the two quantum wells suppresses the inter-layer tunneling and therefore the exciton recombinations, but it is sufficiently thin to provide strong inter-layer Coulomb interaction ~\cite{lozovik}.

Another class of candidate systems consists in Kondo insulators and heavy-fermion materials, which are mixed-valence semiconductors characterized by a flat valence band plus a dispersive conduction band, typically exhibiting strongly correlated behavior. Sham and coworkers~\cite{portengen} have shown that the exciton condensate made of holes and electrons may spontaneously break the lattice inversion symmetry and lead to a ferroelectric phase transition of electronic origin. Such systems are often modeled by the Falicov-Kimball Hamiltonian, which, in its extended incarnation, takes into account the strong inter-band Coulomb interaction and hybridization among them.

The Falicov-Kimball model (FKM) was introduced to explain semiconductor to metal transition and has been extensively used to describe valence transitions in heavy fermion compounds. Its original version contains a dispersive band of itinerant electrons interacting with localized orbitals through local Coulomb interaction. Of late, this has also been used to study excitonic fluctuations in systems like Ta2NiSe5, transition metal dichalcogenides ~\cite{sk} and GdI2~\cite{gdi2_AT} in the limit of small but finite hybridization between the two bands. Possible electronic ferroelectricity has also been studied in FKM~\cite{portengen} where the two participating orbitals have odd and even parity respectively. 

Hybridization between the bands, however, is not the only way to develop excitonic coherence between the two bands. Any dispersion of the localized band can also induce it~\cite{Batista1,Batista2, Sarasua}. A strong hybridization, however, can lead to a gap in the density of states leading to the so called excitonic insulator (EI) phase. If the Coulomb correlation is weak (in comparison to the kinetic energy), the magnitude of the single-particle gap, up to a constant, is given by the excitonic order parameter (EOP). A mean-field description works fairly well in this regime. On the other side, for strong Coulomb interactions, the Mott physics becomes increasingly relevant and the exciton formation and coherence scales move away from each other, the latter goes down with correlation while the former increases. Excitonic fluctuations become increasingly dominant in this regime till the excitonic insulating phase appears through a Mott transition. This transition is beyond the usual treatment of excitons using mean-field theory and calls for methods designed to address strong coupling physics. 

Several numerical schemes, Hartree-Fock (HF) mean-field~\cite{portengen, Farkasovsky, yadav, pradhan1}, Variational Cluster Approximation (VCA)~\cite{Seki}, constrained path Monte-Carlo (CPMC)~\cite{Batista3} and Cluster Perturbation Theory (CPT)~\cite{Senechal} have been brought to bear on this problem in the past and the efforts renewed during the last decade. The mean-field description of the EI phase is similar to the BCS theory of superconductivity and has been worked out long-time ago~\cite{halperin-rice}. Majority of the theoretical considerations on excitons is still based on mean-field theories. 

The advent of dynamical mean-field theory (DMFT) has provided a consistent and capable theoretical framework for strongly correlated electronic systems, and models like the single- and multi-band Hubbard model have been studied with some degree of success using DMFT. However, DMFT (with its various impurity solvers) is essentially based on numerical solutions of relevant impurity models to which the temporal fluctuations of the original correlated models are mapped ~\cite{DMFT}.

A mean-field approach, the slave-rotor mean field theory (SRMF), based on the idea of nominal charge-spin separation provides a fast and efficient means for investigating Mott transition in a variety of situations, particularly in the strong coupling limit. In this approach the Hilbert space of the physical electronic degrees of freedom is enlarged in terms of separate chargon and fermion Hilbert spaces. The unphysical states are then eliminated by enforcing local constraints on the enlarged Hilbert space. In the strong coupling limit, the lattice problem maps on to the problem of interacting slave particles self-consistently coupled to a gauge field. The gauge fluctuations, being weak, provide a framework for studying the Mott-Hubbard physics at intermediate to large coupling~\cite{florens}. In a straightforward extension to a two orbital systems~\cite{lee}, the orbital degrees are also included and the physics of spin, orbital and charge are treated accordingly. 

SRMF has been applied to a variety of strongly correlated electron systems such as the Hubbard model (and its multi-orbital extension~\cite{lee, bierman}) with competing magnetic orders in two-dimensions, Anderson model~\cite{florens}, superconductivity~\cite{paramekanti} and metamagnetism ~\cite{swagata} on bipartite and non-bipartite lattices. This semi-analytical method is useful for discussing correlation effects on the symmetry breaking or single-particle excitation spectra, especially in the insulating state. Owing to disparate charge and spin (or orbital) degrees, it is capable of handling the spin (or orbital)-liquid like states in frustrated systems~\cite{florens,paramekanti}.
The mean-field theory, as discussed above, leads to a growing excitonic amplitude with correlation and misses the  fluctuations at strong coupling~\cite{pradhan1}. We, therefore, take recourse to the slave-rotor mean field approach to investigate the excitons in the extended Falicov-Kimball model. Although a few studies on Hubbard model using SRMF exist, FKM has not been discussed in this context so far. While charge order of various kinds in the pure FKM (without hybridization) has been well studied, excitons in the strong correlation limit of EFKM remains an open issue. 
The outline of this paper is as follows. The extended FKM is introduced in section II. In section III we briefly discuss the slave rotor mean field theory as applied to EFKM. Our results for exciton formation in the weak-coupling as well as strong-coupling regimes are compared with previous weak-coupling results. Sections IV and V discuss various mean field ansatzs for the fermion Hamiltonian and the cluster mean field theory for the rotors. The remainder of the paper deals with discussions and conclusions.
\begin{figure}[!th]
\includegraphics[trim={2cm 0cm 0cm 0cm}, clip, scale=0.18]{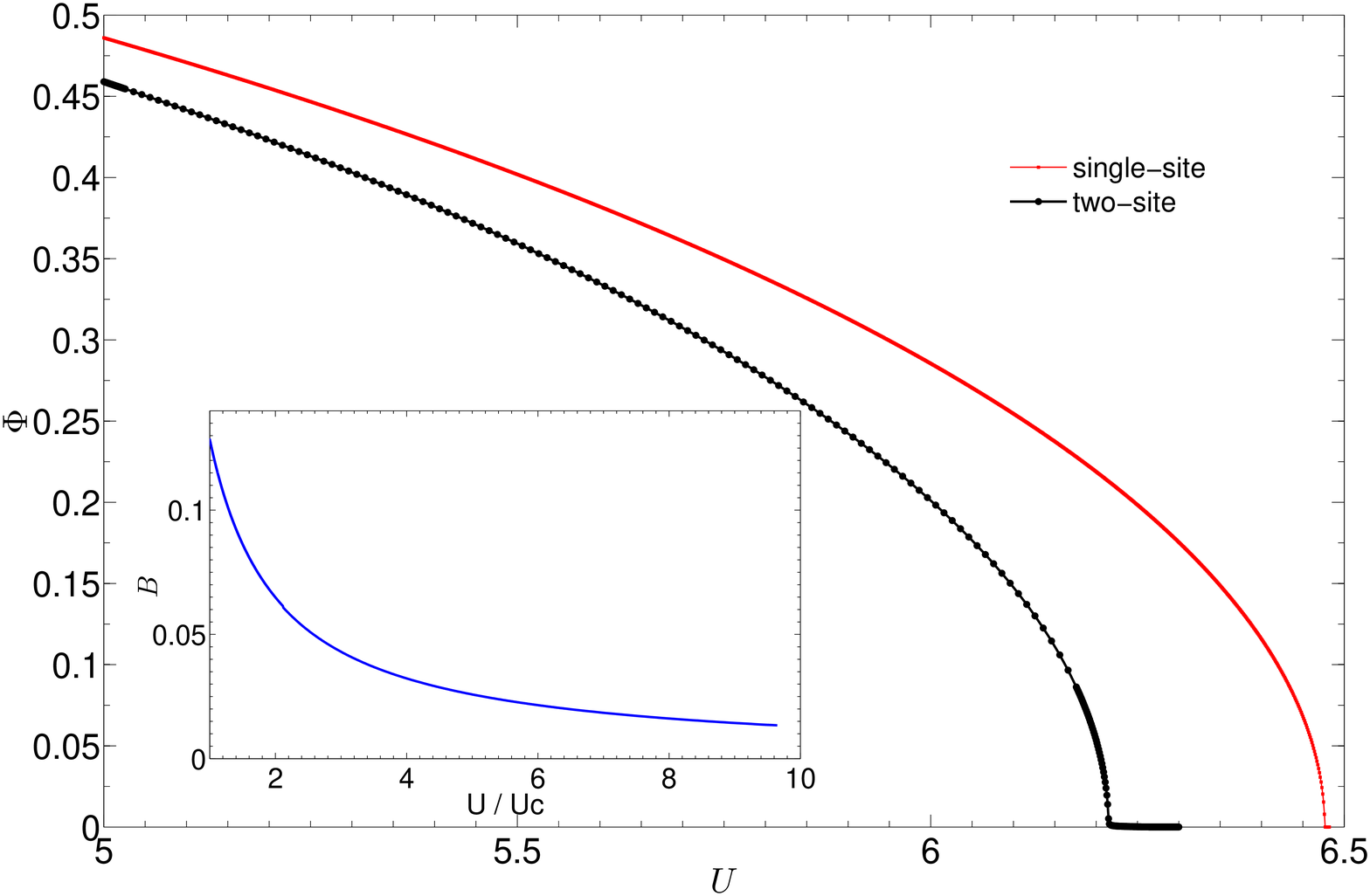}
\caption{(color online) $\Phi$ vs $U/t$ for single-site (red curve) and cluster (black curve) approximation for Hubbard ($t-U$) model. Inset shows the inter-site correlation from the cluster analysis. This curve shows $t^2/U$ scale.} 
\end{figure}

\section{Exciton in a Two-band Model}

The original FKM describes a two-band system of electrons from itinerant (designated $\alpha$) and localized (designated $\beta$) orbitals with local Coulomb interaction $U$ between them. As we are not interested in magnetic properties, we consider a spinless version of it: 

\begin{equation}
\begin{aligned}
H_{FKM} =& - \sum\limits_{< ij >,l} t_{l ij}c_{li}^\dag c_{lj} + U\sum\limits_i c_{\alpha i}^\dag c_{\alpha i}c_{\beta i}^\dag c_{\beta i}+ \hfill \\
\sum\limits_{i, l} \mu_{l}c_{li}^\dag c_{li}
\end{aligned}
\end{equation}

\noindent  where sum over $l$ runs through orbitals $ \alpha,\, \beta$. If $t_{\beta} = 0$, the above Hamiltonian corresponds to (spinless) FKM. Here, ${<i,j>}$ are nearest-neighbour site indices on a square lattice (lattice constant  = 1), $c_{\alpha i}$ ($c_{\beta i}$) are itinerant (localized) electron annihilation operators at site $i$. The first term is the kinetic energy due to hopping between nearest neighbors, where, $t_{\alpha} \, (t_{\beta})$ is the hopping integral for the electrons in the itinerant (localized) band. $t_{\alpha}$ is kept fixed at 1 throughout our calculation as the scale of energy. The second term represents the on-site Coulomb interaction between electrons in bands 1 and 2. This Hamiltonian commutes with $\hat{n}_{2i}$, in which case local occupancy of band-$2$ electron is either 0 or 1. This renders the model `solvable', albeit numerically, by annealing over the localized electron configurations~\cite{pradhan1}. No coherence between electrons from the two bands is possible in this situation. If hybridization $\sum\limits_i V(c_{\alpha i} ^\dag c_{\beta i} + h.c.)$ between these two bands is included, the local $U(1)$ symmetry of the $\beta$ electrons is lifted. The same happens for an extended FKM with finite bandwidth $\sum\limits_{< ij >} -t_{\beta} c_{\beta i}^\dag c_{\beta j} $. So the Hamiltonian now reads, 

\begin{equation}
\begin{aligned}
H =& -\sum\limits_{\langle ij \rangle }t_{\alpha}c_{\alpha i}^\dag c_{\alpha j}
-\sum\limits_{\langle ij \rangle}t_{\beta}c_{\beta i}^\dag c_{\beta j} \nonumber\\&+\sum\limits_{li} E_l c_{li}^\dag c_{li}\nonumber\\& + U\sum\limits_i \hat n_{\alpha i}\hat n_{\beta i} + \sum\limits_i V(c_{\alpha i} ^\dag c_{\beta i} +h.c.).
\end{aligned}
\end{equation}

\noindent where $\hat n_{\alpha i} = c_{\alpha i}^\dagger c_{\alpha i}$ and similarly for $\hat n_{\beta i}$. In the limit $t_{\alpha} = t_{\beta}$, the above model is the Hubbard model in the pseudospin (orbital) space. However, if $\alpha$, $\beta$ electrons have different hopping amplitudes, i.e.,  the two pseudospin degrees admit of different dispersions, then we have a more general model. In the limit, when $\beta$ orbital becomes dispersionless the $\beta$-electron density is locally conserved due to local $U(1)$ symmetry and excitonic order at finite temperature is prohibited (Elitzur Theorem). The inter-orbital hybridization between the two bands (transverse field in the Hubbard model Eqn.(1)), however, breaks the symmetry making such an order possible. 

\section{Formulation}
In this section, we briefly review the formulation of the SRMF for this two-orbital model~\cite{lee}. The electronic Hilbert space at a single lattice site has four states: $|0 \rangle$, $|\alpha\rangle$, $|\beta\rangle$, $|\alpha \beta\rangle$. In the slave rotor (SR) representation, the electronic charge and the orbital degrees are described by a charged rotor and a fermion carrying the orbital index respectively. The direct product space of the rotor and fermion states contains the four physical states at every site, with additional, unphysical states in an enlarged Hilbert space which is mapped back on the physical one via a local constraint acting like a gauge term. In the SR representation, the electron number is equal to the fermion number. The electron annihilation operators, for example, are written in the SR representation as
\begin{figure}[!th]
\includegraphics[trim={0.5cm 0.3cm 0cm 0.3cm}, clip, scale=0.15]{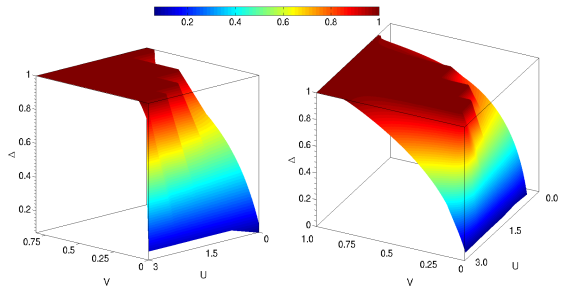}
\caption{\label{Fig.2} (color online) ${\Delta}-U-V$ phase diagram: a slave-rotor mean field behavior. The left Figure shows results for single-site case and right figure corresponds to two-site behavior.} 
\end{figure}

\begin{eqnarray}
c_{\alpha i} \equiv f_{\alpha i}\exp({i}{\theta_i}); \,\,
c_{\beta i} \equiv  f_{\beta i}\exp({i}{\theta_i}).
\end{eqnarray}
In terms of fermions $f_{\alpha i},\,f_{\beta i}$ and rotor  $\exp (\pm {i\theta}$) operators, this model can be written as (see Florens, et al.~\cite{florens} for details)

\begin{equation}
\begin{aligned}
H_{SR} =&-\sum_{<ij>l}t_{ijl} f_{li}^{\dag}f_{lj}e^{-i\theta_i}e^{i\theta_j}+
\frac{U}{2}\sum_{i} n_{i}^{\theta}(n_{i}^{\theta}-1)\nonumber\\& + V\sum_{i} f_{\alpha i}^\dag \exp(-i\theta_{i})f_{\beta i}\exp(i\theta_i)\nonumber\\& + \sum_{i,l}\mu_{fl} n_{li}^{f}
\end{aligned}
\end{equation} 
In the single-site SRMF theory, the Hamiltonian for the rotor and auxiliary fermion sectors are decoupled as: 
\begin{equation}
\begin{aligned}
H_f=&-\sum_{<ij>l} t_{lij}B_{ij}f_{li}^\dag f_{lj}  + \mu_f \sum_{li}n_{li}^f 
\nonumber\\&+ V\sum_{i}(f_{\alpha i}^{\dag}f_{\beta i}+ h.c.)
\end{aligned}
\end{equation}  
\noindent and
\begin{eqnarray}
H_\theta=&-2\sum_{<ij>l}t_{lij}\chi_{lij}e^{-i\theta_{i}}e^{i\theta_{j}} + U/2\sum_{i}(n_{i}^{\theta})^{2}-\mu_{\theta}\sum_{i}n_{i}^{\theta}
\end{eqnarray}
\noindent where, for single-site approximation, $B_{ij} = \langle e^{-i\theta_{i}}e^{i\theta_{j}}\rangle_{\theta}$, $B_{ij} = \langle e^{-i\theta_{i}}\rangle e^{i\theta_{j}}$,\,\, $B = \phi^{2}$, $\chi_{lij} = \langle f_{li}^{\dag}f_{lj}\rangle_{f}, \,\, \mu_{f},\,\mu_{\theta}$ are chemical potentials used to control $\langle n_{i}^{\theta}\rangle$ and $\langle n_{i}^{f}\rangle$ for the number constraint: $\langle \sum_{l}n_{li}^{f}\rangle + \langle n_{i}^{\theta}\rangle = 1$. The Hamiltonian for the rotor part is diagonalized numerically, $\phi$ and $B_{ij}$ are calculated from the ground state and fed back to $H_f$ to get  $\chi$. This is then put in $H_\theta$ and the process repeated till convergence in $\chi$, $\phi$ and $B_{ij}$ is reached. In a homogeneous situation, these site and bond order parameters become independent of the site indices and the site indices are henceforth dropped.
\subsection{Single-site analysis}

The rotor kinetic energy acts as the quasiparticle (QP) weight ($Z$) for the fermions. If $\phi^{2}$ vanishes, charge fluctuation is quenched and the system is an insulator. The single-site rotor Hamiltonian is now 

\begin{eqnarray}
H_{\theta}=-4(t_{\alpha}\chi_{\alpha} + t_{\beta}\chi_{\beta})\phi(e^{-i\theta}+e^{i\theta})+U/2(n^{\theta})^{2}-\mu_{\theta}n^{\theta}
\end{eqnarray}

\noindent The order parameter for electron-hole bound state is defined as,
\begin{eqnarray}
\Delta= \langle f_{\alpha i}^\dag f_{\beta i}\rangle
\end{eqnarray}

In this way, the rotor sector $H_\theta$ and free-fermion sector $H_f$ are coupled and the rotor Hamiltonian is solved on a finite cluster self-consistently coupled to an order parameter bath (rest of the lattice). However, the single-site theory has its limitations. It fails to identify the long range correlation (the pseudospin exchange) and underestimates the Mott scale. It is very similar to the Gutzwiller approximation in that the double occupancies are completely eliminated at the Mott transition. A minimum cluster of two sites is then required to incorporate these. 

\subsection{cluster Analysis}

We need to extend the theory from site to cluster to include intersite correlations. The rotor Hamiltonian for the cluster SRMF is 
\begin{align}
H_{\theta}=&-\sum_{l}t_{l}\chi_{l}(e^{-i\theta_{1}}e^{i\theta_{2}}+h.c.)
\nonumber\\&+\sum_{l}3t_{l}\chi_{l}(e^{-i\theta_{1}}+e^{-i\theta_{2}}+h.c.)
\nonumber\\&+U/2(n_{1}^{\theta})^{2}+U/2(n_{2}^{\theta})^{2}-\mu_{\theta}
(n_{1}^{\theta}+n_{2}^{\theta})
\end{align}

\noindent For a two-site extension, the intra-site correlation is accounted, considering a bond connecting these two-sites; along with the site interaction terms. This cluster mean-field Hamiltonian is again diagonalized numerically to obtain the eigenvalues and the ground state wave function of the rotor part in a two-site basis $|n_1^{\theta}>,|n_2^{\theta}>$, where, $B^{'}=\Phi^{2}$ with $\Phi=\langle e^{\pm i\theta}\rangle$ and $B$ = $\langle e^{-i\theta_{1}}e^{i\theta_{2}}\rangle$. When $\Phi$ goes to $0$ (i.e., the insulating phase), unlike the single-site case, the nearest neighbor inter-site correlation could assume a non-vanishing value in the cluster approximation. The first term in the above equation gives a finite rotor kinetic energy $\sim 1/U$. A bond approximation approach, therefore recovers the inter-site exchange correlation ($\sim t^{2}/U$). The cluster mean field theory focuses on a finite cluster of sites and treats the influence of the sites outside (the ``bath") via a mean field order parameter. A larger cluster size yields much better results for $\Phi$ and $B_{ij}$. There are qualitative difference between the results obtained from the single-site mean field theory and the cluster SRMF: in single site theory, the rotor kinetic energy $B_{ij}$ is the square of the order parameter $\Phi^2$. When it vanishes, Mott insulating phase sets in. This amounts to neglecting all density fluctuations within the Mott phase, too crude an approximation close to the Mott transition. By contrast, the cluster mean field theory captures the short distance correlation. The main advantage of the cluster mean field theory is that the short range correlation functions of the rotor are properly taken care of. In contrast to Hartree-Fock Mean-Field theory, SRMF provides an exact treatment of the quartic interactions in the above Hamiltonian.

\section{Results}
\subsection{Symmetric Case ($t_{\alpha} = t_{\beta}$)}
\noindent  When $t_{\alpha} = t_{\beta} = t$ and $V = 0$, the model is the Hubbard model, which shows a metal-insulator transition (MIT) driven by local correlation. We reproduce the  single site and cluster results for this model for consistency. The nonlocal correlations are accounted for in the two-site cluster, the critical $U/t$, at which metal-insulator transition occurs, is now lower. The metallic phase disappears through a Brinkman-Rice transition~\cite{Rice}, at which the quasiparticle (QP) weight ($Z = \Phi^2$) vanishes and the effective mass diverges. It preserves Hubbard bands in the insulator, and a “preformed” Mott spectral gap opens up discontinuously at the transition is found, as in DMFT~\cite{DMFT}. The critical values of $U/t$, at which metal-insulator transition (MIT) occurs is $6.483t$ and $6.219t$ (in the absence of hybridization) for single and two site cases respectively, these values of $U_{c}$ match well with earlier results~\cite{swagata}. The parameter $B$ (inset, Fig.1) signifies that the non-local fluctuation remains finite even in the insulating phase and approaches zero in the $U/t \rightarrow \infty$ limit. 

\begin{figure}[!th]
\includegraphics[trim={0cm 0cm 0.0cm 0cm}, clip, scale=0.18]{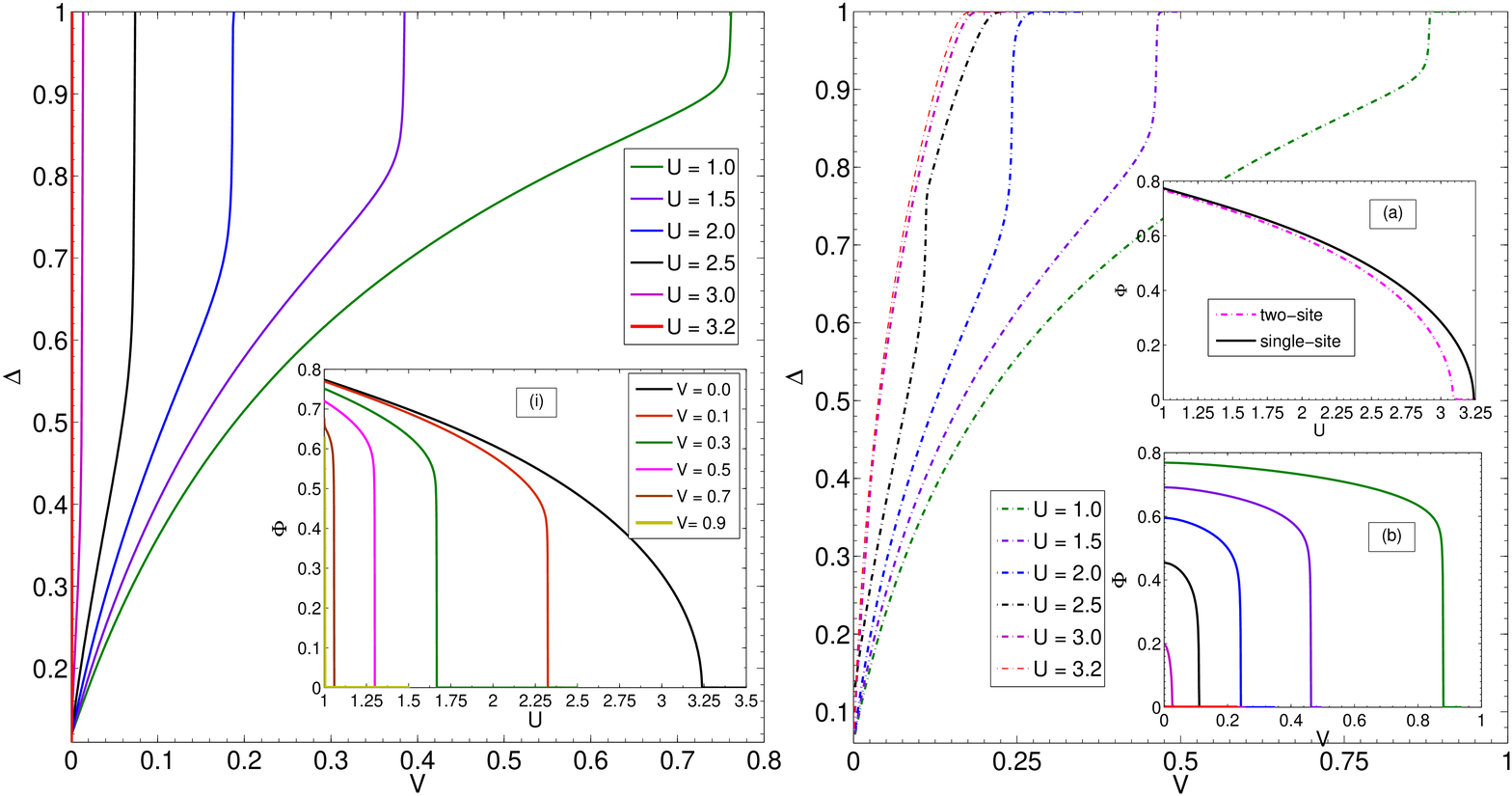}
\caption{\label{Fig.3}(color online) Mean-field results for exciton order parameter as a function of hybridization strength $V$ for different sets of $U$ for single site (left panel) as well as two-site case (right panel). In the left panel, the order parameter rises from zero steeply with $V$ for $V < V_{c}$. Inset shows the variation of $\Phi$ with $U$ for different $V$. The critical $U_c$ moves towards left as we increase $V$. Right panel shows the variation of EOP with $V$ with bond-approximation. Right panel inset (a) shows the critical $U$ for single and two-site approximation is 3.21t and 3.105t respectively. Inset (b) is the plot for rotor kinetic energy $(\Phi)$ with $U$ for different $V$. The effect of cluster approximation is strong for large $U$-regime.} 
\end{figure} 

\subsection{Exciton in an Extended Falicov Kimball Model}
\begin{figure}[!th]
\includegraphics[trim={0.0cm 0cm 0.0cm 0cm}, clip, scale=0.29]{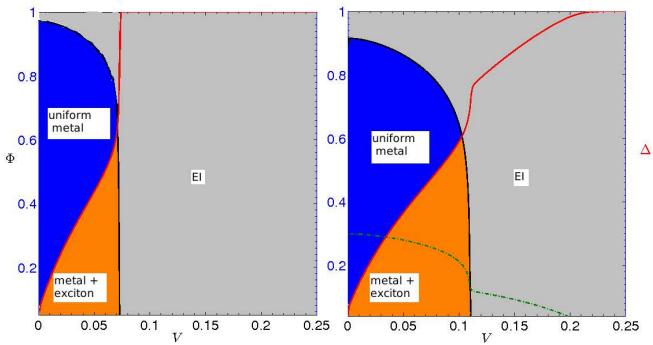}
\caption{\label{Fig.4} Phase diagram showing EOP with hybridization strength $V$ for $U = 2.5$. Left panel shows single site results, while right panel for cluster analysis. The metallic and the metal + exciton regime are enhanced in cluster approximation. The transition from the metal + exciton mixed state to excitonic insulator (EI) is first order for single site case. EOP takes over exactly when $\Phi$ vanishes  for a single site case. But in two-site approximation, the transition is weakly first order (at $U = 2.5$) as one incorporates the inter-site correlation $B_{ij}$; $\Delta$ has finite value even when $\Phi = 0$ and saturates as $B_{ij}$ goes to zero (shown by dashed green line).} 
\end{figure} 
When one of the two-spin components in the Hubbard model goes to zero, it is the Falicov Kimball model. We check the critical $U$ for this model at which the rotor K.E. $\phi$ vanishes; suggesting a metal-insulator transition, in single as well as in two-site approximation. It is found that as $t_{\beta} = 0$, the required $U (=U_c)$ at which $\Phi$ goes to zero just becomes half of $U_c$ for the Hubbard model. Fig.3 and its inset show that the critical $U$ for single site and two-site are $3.24t$ and $3.105t$ respectively. 
If the hybridization term $H_v = {{\sum\limits_i {V({c_{\alpha i}}} ^\dag }{c_{\beta i}} + h.c.)}$ between these two bands are included in the Hamiltonian, the local $U(1)$ symmetry of the $\beta$-electrons is lifted. The same happens for an extended FKM with finite $\beta$-electron bandwidth (${\sum\limits_{ < ij > } {{-t_{\beta}}(c_{\beta i}}} ^\dag c_{\beta j} + h.c.)$).

Fig.2 shows a 3D plot of excitonic order in the $U-V$-plane for single as well as two-site extension. The mean-filed behavior has salient differences for site and cluster theory. In the low-$U$ limit, both single and two-site results show a continuous change in the order parameter. This regime can be compared to earlier results obtained using Hartree-Fock mean field ~\cite{pradhan1}. Here $\Delta$ increases with $V$ and $U$ and saturates after a certain $V$. To see the difference between Hartree-Fock mean-field theory in the weak-coupling regime, and the treatment using SRMF, one compares the variation of $\Delta$ in the $U-V$ plane (Fig.2) with earlier study \cite{pradhan2}. The jump in EOP seen in SRMF is absent in the HF mean-field. The large-$U$ regime, better captured in SRMF theory, shows deviations from HF as well as between single and cluster approximations. In the single-site case, with increasing $U$, excitonic order shows a first order jump and goes to its saturation value quickly with increasing $U$. On the other hand, two-site results show a transition from weakly first order back again to a second order transition with $U$.

Fig.3 corroborates the same as shown in Fig.2 in detail. It shows the variation of excitonic order parameter with hybridization strength $V$; as $U$ increases the critical value of $V$ at which excitonic insulating state appears is  shifted towards left. Similarly, an increase in $V$ also leads to MIT at a critical $V_c$ which goes down as $U$ increases (inset, Fig.3). For small values of $U$, it is obvious that a hybridization will not change the low-energy nature of the metallic phase qualitatively. The rotor kinetic energy $\Phi$ vanishes at a small $U$ as we increase $V$, which is again associated with the electron-hole pair (exciton) formation and condensation. For a large $U$, the gap is robust and less affected by the hybridization $V$. A large $V$ also makes the system gapped and the value of the gap is of the order of $2V$; giving rise to a metal-insulator transition. Expectedly, at a critical $V$ ($V_c$), the EOP saturates with a first order jump. Well below $V_c$, there is a co-existance of metallic as well as excitonic phases; where at $V_c$ the system goes over to an excitonic insulator phase with a first order jump in the EOP.
\begin{figure}[!th]
\includegraphics[trim={1.5cm 0cm 0.0cm 0cm}, clip, scale=0.182]{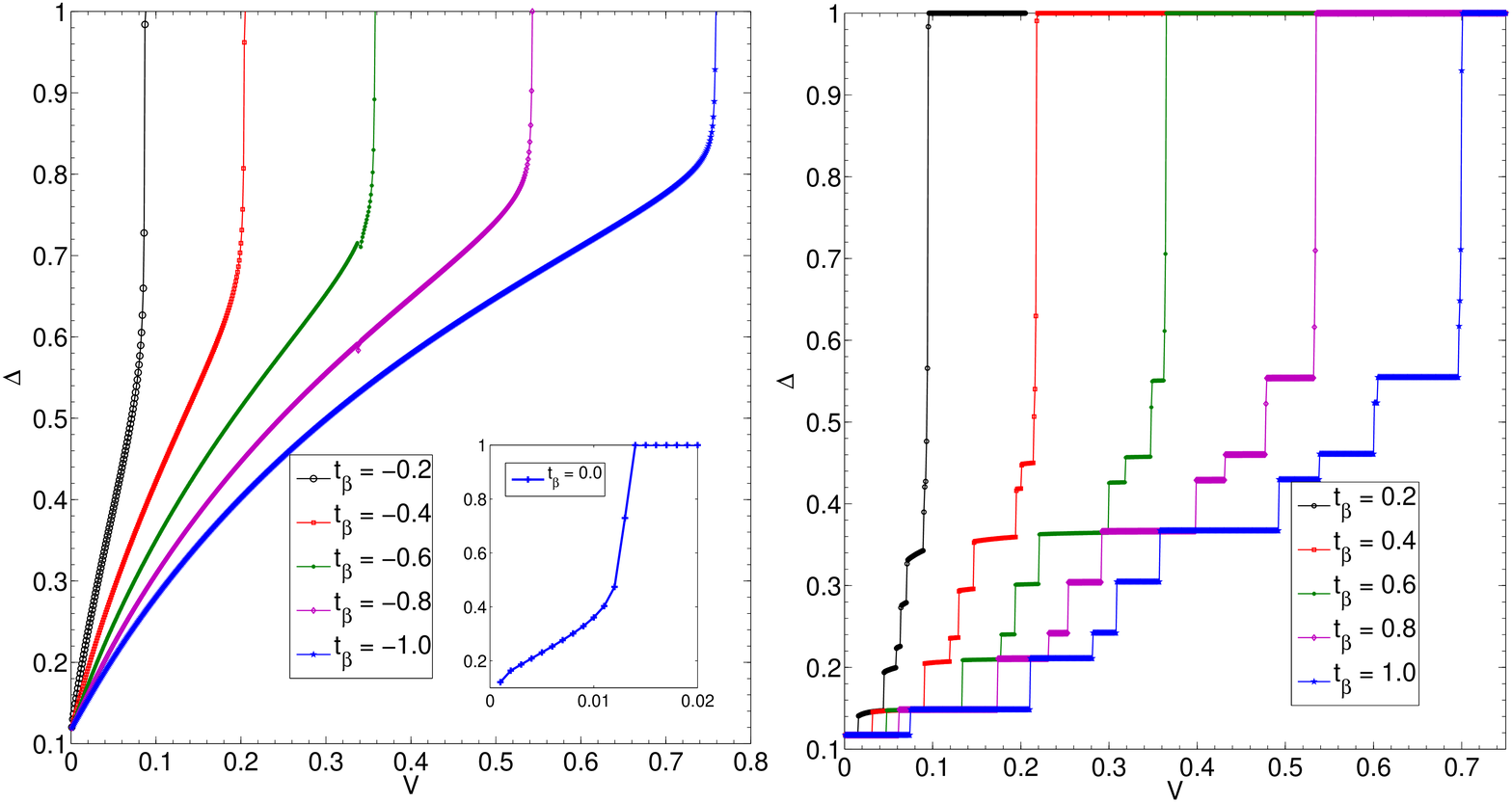}
\caption{\label{Fig.5} The variation of EOP $\Delta$ when $t_{\alpha}/t_{\beta}> 0$ (left panel) and $t_{\alpha}/t_{\beta}< 0$ (right panel) with $V$ for different $t_{\beta}$ (single site result). On-site interaction $U$ is kept fixed to 3.0. Inset shows the variation of $\Delta$ with $V$ when $t_{\beta} = 0$.} 
\end{figure} 

The phase diagram in Fig.4 details the nature of phase transition in single as well as in two-site case. The transition from the metal-exciton mixed state to excitonic insulator (EI) is of first order in nature. In the two-site case, even when $\Phi \sim 0$, $\Delta$ does not reach the saturation value, still varies with $V$ and reaches a saturation value when $B_{ij}$ goes to zero. The (green) dashed line in Fig. 4 shows $B_{ij}$. There are several differences between the earlier BCS mean-field result. The exciton formation leads to the insulating state in the earlier treatment while a co-exiting excitonic metal state appears in the SRMF. In addition, the non-zero $B_{ij}$ beyond $\phi =0$ in here signifies a region analogous to the spin-liquid in the Hubbard model~\cite{florens}. In the bond approximation a new scale $t^2/U$ emerges signifying the appearance of long-range inter-orbital excitonic fluctuations, analogous with the anti ferromagnetic exchange in the Hubbard model at half-filling~\cite{swagata}.  

\begin{figure}[!th]
\includegraphics[trim={1.0cm 0cm 0.0cm 0cm}, clip, scale=0.18]{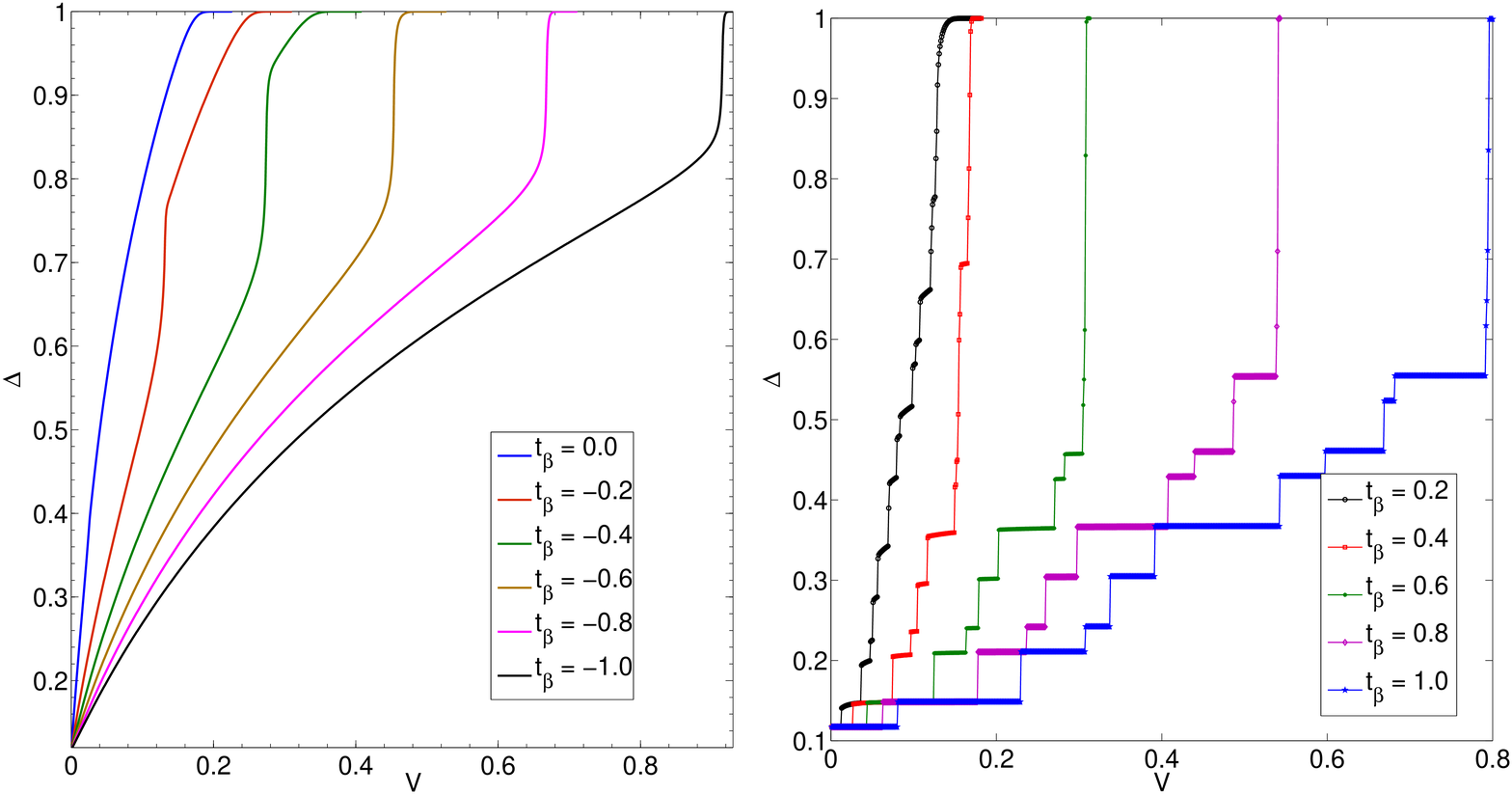}
\caption{\label{Fig.6} The variation of EOP $\Delta$ when $t_{\beta}/t_{\alpha}< 0$ (left panel), $t_{\beta}/t_{\alpha}> 0$ (right panel) with $V$ for different $t_{\beta}$ for two-site approximation. On-site interaction $U$ is kept fixed at 3.0.} 
\end{figure} 
\subsection{Effect of finite $t_{\beta}$}

For an asysmmetric hopping between $\alpha$ and $\beta$, an increase in $t_{\beta}$ fermion kinetic energy increases hence the critical $U$ for metal-insulator transition increases. On the other hand, the excitonic order parameter goes down as the probability of formation of electron-hole bound state reduces with increased ${\beta}$-electron hopping. In addition to that, steps appear (absent for $t_{\beta}= 0$) in the $\Delta$ vs $V$ curves and the step-size increases with the value of $t_{\beta}$.

\noindent Steps appear (see Fig.5 (right panel), Fig.6 and Fig.7) when both the spins has same parity ($t_{\beta}/t_{\alpha} > 0$), no steps for the opposite parity ($t_{\beta}/t_{\alpha} < 0$) case. The step-size increases if we incorporate inter-site correlation. In the two-site approximation, the effect of hopping becomes interesting. When $U \simeq U_c$, $t_{\beta} = 0$ (FKM limit), it is found that the $\Delta-V$ curve is almost continuous, but as we see from Fig. 6, the order parameter $\Delta$ becomes first order with the inclusion of $t_{\beta}$. One can go from weakly second order to first order by tuning $t_\beta$.
The excitonic order always reduces with $t_{\beta}$, the probability of formation of excitonic bound state becomes less and less. Fig.7 shows how the exctonic order reduces with increasing $t_\beta$. 

In the strong-coupling regime, the EFKM ($V \ne 0$) can be mapped onto the spin-1/2 Ising XXZ model with a transverse magnetic field. In that case, the spontaneous EI order corresponds to the spontaneous magnetization in the XY plane. Therefore, the excitonic order in EFKM is like the metamagnetism in the half-filled Hubbard model: the variation in the magnetization with external applied magnetic field. This is reflected in $\Delta-V$ behavior. The Zeeman term in the metamagnetism in Hubbard model couples to the z-component of  spin, while $V$ term in the EFKM is like a transverse field. 
In the absence of $U$, the system can also open a gap which is driven by $V$; known as excitonic gap which is exactly equal to $2\Delta$, i.e., twice the excitonic order parameter. At low $U$, large $V$ is required to make $\Phi$ zero, consequently one enters an insulating phase. This is a band-insulator driven by $V$. On the opposite, large $U$ limit, even a small $V$ opens a gap of the order $V$. Here a correlated insulator is realized.

\begin{figure}[!th]
\includegraphics[trim={0.0cm 0cm 0.0cm 0cm}, clip, scale=0.18]{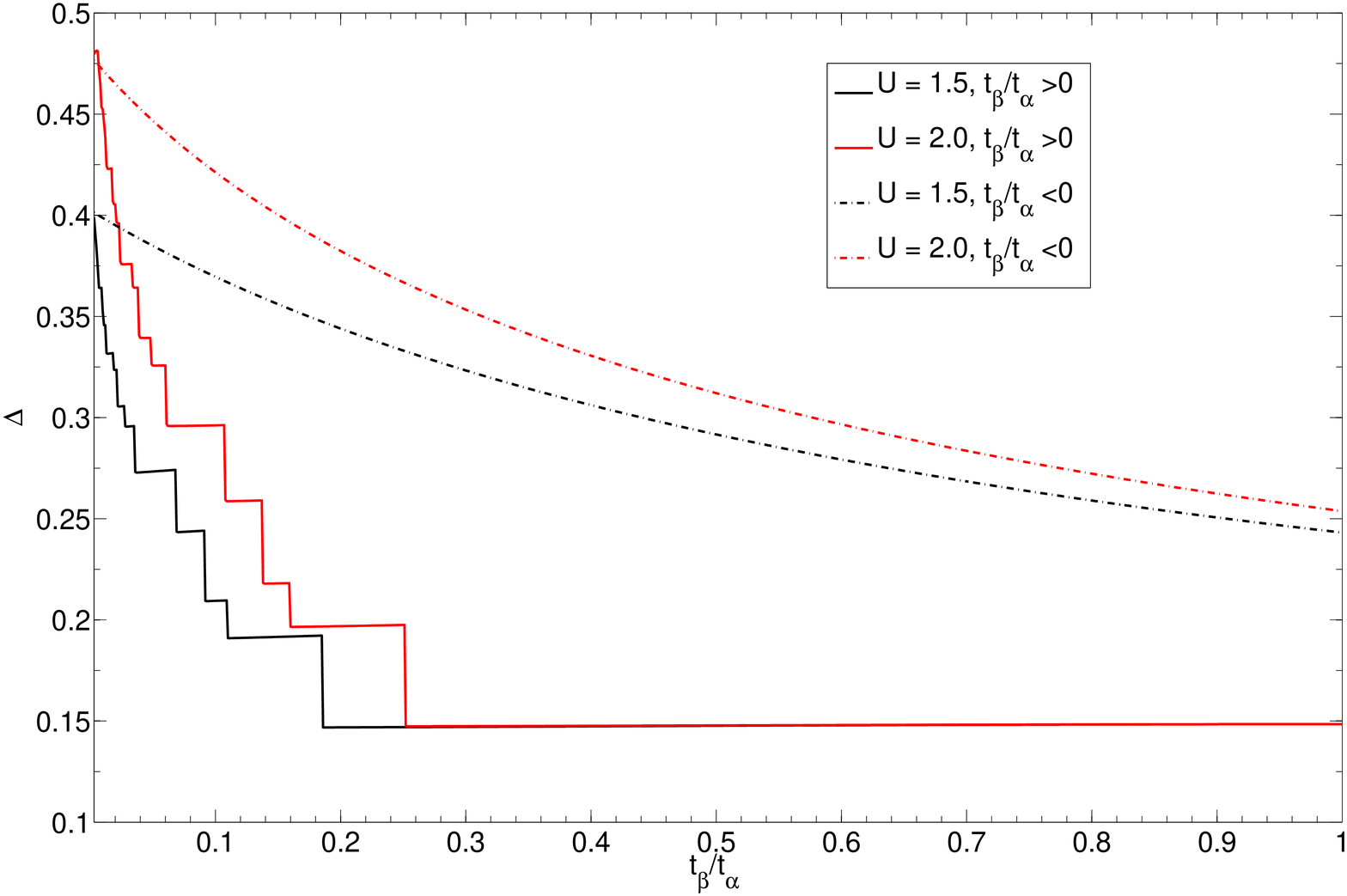}
\caption{\label{Fig.7} The variation of EOP $(\Delta)$ with $t_\beta$ for $V = 0.1$ for two cases when $t_{\beta}/t_{\alpha
}>0$ and $t_{\beta}/t_{\alpha}<0$.} 
\end{figure} 
\section{CONCLUSIONS}

In a slave-rotor formalism, we unravel the excitonic physics in a two-band system in the strongly correlated regime. The inclusion of hybridization among these two levels gives rise to an electron-hole bound state defined by an order parameter $\Delta$; which enhances with $V$ for all $U$-regime. Metal-insulator transition can be tuned by $V$ also, it is found that for low $U$, a large $V$ is required to make the rotor kinetic energy zero. The critical $U$ for MIT moves towards lower values with increasing $V$. After a certain critical correlation $U_c$ or $V_c$ there is a first-order jump in the excitonic order parameter associated with an MIT. The cluster analysis gives $t^2/U$-scale in the calculation and correctly describes the metallic and insulating phases. There is coexistence of metallic and excitonic phases and after a certain hybridization, electrons are completely localized and an EI state follows. The effect of $t_{\beta}$ is also interesting, the behavior of $\Delta$ depends on the sign of $t_{\beta}$; steps appear when $t_{\beta}/t_{\alpha} > 0 $. The excitonic order parameter reduces with increasing $t_{\beta}$.

\acknowledgements
We acknowledge S Acharya and N Pakhira for useful discussions regarding the formulation of SRMF technique. Centre for Theoretical Studies, IIT Kharagpur is acknowledged for providing computer facilities.

\begin{figure}[!th]
\includegraphics[trim={0.0cm 0cm 0.0cm 0cm}, clip, scale=0.18]{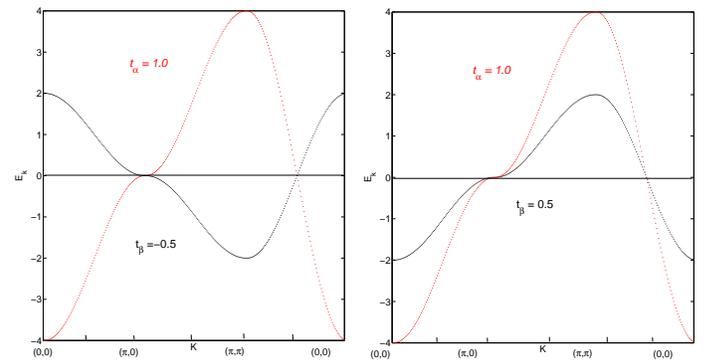}
\caption{\label{Fig.8} Noninteracting tight-binding band structure of the $\alpha$-orbital (red line) and $\beta$-orbital (black-dotted
line).} 
\end{figure} 

\begin{figure}[!th]
\includegraphics[trim={0.0cm 0cm 0.0cm 0cm}, clip, scale=0.18]{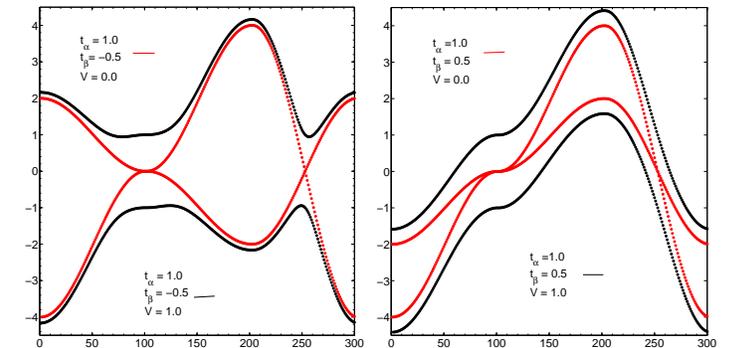}
\caption{\label{Fig.9} Excitonc band insulator.} 
\end{figure} 

\end{document}